# Study of Electro-Caloric Effect in Ca and Sn co-Doped $BaTiO_3$ Ceramics

Sanjay Kumar Upadhyay[1,2*], Iram Fatima [3] and V. Raghavendra Reddy[1]

[1]*UGC DAE Consortium for Scientific Research, Khandwa Road, Indore-452001, India*
[2]*Tata Institute of Fundamental Research,Mumbai-400005, India*
[3]*School of Physics, DAVV Indore-452001, India*

**E-mail: skuphysics@gmail.com*

**Abstract**

The present work deals with the study of structural, ferroelectric, dielectric and electro-caloric effects in lead free ferroelectric polycrystalline $Ba_{1-x}Ca_xTi_{0.95}Sn_{0.05}O_3$ (x= 2, 5 and 10 %) i.e., Ca, Sn co-doped $BaTiO_3$ (BTO). Phase purity of the samples is confirmed from X-ray data by using Rietveld refinement. $^{119}$Sn Mössbauer reveals homogenous phase as well as iso-valent substitution of Sn at Ti site. Enhancements in ferroelectric and dielectric properties have been observed. Indirect method which is based on Maxwell equation has been used to determine the electro-caloric (EC) effect in the studied ferroelectric ceramics and maximum EC coefficient is observed for $Ba_{0.95}Ca_{0.05}Ti_{0.95}Sn_{0.05}O_3$.

## 1. Introduction

In recent years, it is speculated that electro-caloric (EC) effect exhibited by ferroelectric materials are potential candidates for future generation cooling devices as compared to mostly studied magneto-caloric materials [1]. $BaTiO_3$ (BTO) based compounds are considered to be attractive materials toward this aspects compared to lead based ferroelectric materials in terms of the environment friendly and also their stability over the broad temperature range [2, 3, 4]. In BTO based systems by suitable doping, one can tune various parameters like transition temperature ($T_C$), dielectric constant, electro-striction, piezoelectric coefficient, ferroelectricity etc [5]. It is to be noted that doping either at Ba or at Ti site affects the $T_C$ and offer the possibility of tuning $T_C$ close to room temperature [6, 7]. This aspect is being utilized by many research groups to enhance the EC effect in FE materials [8, 9]. Since close to $T_C$ as the dielectric susceptibility diverges, one would expect the enhanced EC effects close to $T_C$ and therefore in recent years enhanced EC effects are reported near phase transition temperature as compared to other temperature regions [8, 9, 10, 11]. For example, Singh et al, has reported the EC effect in $Ba_{0.92}Ca_{0.08}Zr_{0.05}Ti_{0.95}O_3$ (dT/dE= 0.253 K mm/kV) at its ferroelectric to paraelectric transition temperature of 403 K [8]. Wang et al., studied $Ba_{0.94}Ca_{0.06}Ti_{1-x}Sn_xO_3$ (x= 5 to 20%) ceramics from EC effect point of view and reported an EC effect of about 0.4 K mm/kV for x= 12.5% sample near room temperature with an applied field of 6 kV/cm [9]. Recently, we have studied that Sn doping at Ti site of BTO results in tuning one of the phase transition temperatures (between orthorhombic to tetragonal) close to room temperature and also in abnormal grain growth with better FE properties [12]. As a consequence of this, better EC properties (~0.2 K mm/kV) at room temperature with applied electric field as low as 3.6 kV/cm are observed for 5% Sn doping at Ti site [12]. Starting from this optimized composition of BTO, we have further attempted to improve the EC coefficient by doping at the A-site of BTO i.e., with Ca doping at Ba site. In the present work we report the effects, of co-doping in BTO with Ca at Ba site and 5% Sn at Ti site, on the EC behavior. All the studied ceramics are characterized by X-ray diffraction, $^{119}$Sn Mössbauer, scanning electron microscopy (SEM), PE loop and impedance analyzer spectroscopy. It is observed that 5% Ca doping at Ba site in $BaTi_{0.95}Sn_{0.5}O_3$ resulted in better EC value (of about 0.5 K.mm/kV) as compared to parent compound.

## 2. Experimental details

Conventional solid-state route has been used for the preparation of polycrystalline $Ba_{1-x}Ca_xTi_{0.95}Sn_{0.05}O_3$ (x= 2, 5 and 10 %) samples. $BaCO_3$, $TiO_2$, $SnO_2$ and $CaCO_3$ are used as an initial precursors for the synthesis of the said compound. The final sintering has been done at 1300 $^0$C for 2 hours. Samples are accordingly designated as Ca2, Ca5 and Ca10 for x=2, 5 and 10 % (representing the Ca content in $Ba_{1-x}Ca_xTi_{0.95}Sn_{0.05}O_3$) respectively. SEM measurements (model LEO s-440i) are carried out to study the microstructure of the prepared samples. Standard PC-based Mössbauer spectrometer equipped with a WisEl velocity drive in constant acceleration mode has been used for the $^{119}$Sn Mössbauer measurements. The spectrometer is calibrated with natural iron and the reported isomer shift values are with respect to $SnO_2$. The ferroelectric measurements (P-E loop) are carried out by P-E loop tracer supplied by M/s Radiant Instruments, USA. Temperature dependent PE loops were carried out using home-made sample holder. Commercial Physical Properties Measurements System (Quantum Design, USA) was used to measure complex dielectric permittivity using an Agilent E4980A LCR meter with a home-made sample holder with several frequencies (1 kHz to 100 kHz) and with a bias voltage of 1 V.

## 3. Result and discussion

Figure 1 represents the X-ray diffraction data of all the studied polycrystalline viz., Ca2, Ca5 and Ca10 samples. The obtained data are further fitted with the Rietveld refinement by using FullProf program [13]. The refinement confirms the tetragonal structure (*P4mm* space group) of all samples at room temperature. The lattice parameters obtained after the refinement are: a=3.997(2) Ǻ, c= 4.020(1) Ǻ for Ca2, a=3.995(1) Ǻ, c= 4.021(1) Ǻ for Ca5 and a=3.988 (1) Ǻ, c= 4.014 (2) Ǻ for Ca10. It may be noted that we have also tried for the higher composition of 15 % Ca at Ba site, but a small impurities phase has been found. For understanding the microstructure and effect of doping, SEM measurements are carried out. From SEM image it has been observed that there are significant differences in the microstructure of the studied samples with doping. The average grain size for x=2%, 5% and 10% samples are 1, 7.5 and 5µm respectively and presented in figure 2.

Figure 3 shows the $^{119}Sn$ Mössbauer spectra at room temperature for the $Ba_{1-x}Ca_xTi_{0.95}Sn_{0.05}O_3$ (x=2, 5 and 10%) sample. Mössbauer measurement is one of the best techniques for the confirmation of the oxidation state of the Mössbauer active atom. It is known that Sn can exit in two stable oxidation ($Sn^{+2}$ and $Sn^{+4}$) state. However, the observed isomer shift of the all samples is found to close to zero conclusively indicates the presence of $Sn^{4+}$ in all the studied samples, i.e. iso-valent substitution of Sn at Ti site. The observed symmetric line shape for all the samples also rules out the possibilities of segregation of Sn.

Figure 4 shows the room temperature ferroelectric (P-E) data of all the studied compounds measured at 50 Hz. For Ca2, Ca5 and Ca10 the maximum polarization ($P_{max}$) value are 6.44, 7.76, 3.71 µC/cm$^2$, the remnant polarization value are 2.50, 2.62 and 1.28m µC/cm$^2$ and the coercive field are 2.36, 1.7 and 1.9 kV/cm respectively. The results established that there is change in the polarization value with the Ca addition at Ba site. Figure-5 shows the dielectric data of all studied compounds in the temperature regime of 325-10 K with the frequency of 1 kHz-1MHz. The data was measured in the cooling sequence. Due to the limitation on the higher temperature side, we could not capture the cubic-tetragonal transition in the samples from the dielectric data. But the cubic to tetragonal transition is observed from high temperature P-E data (figure-6, 7). The other observation is that Ca addition contributed to the increase of the maximum applied field on the studied compound (from 8 to 12 kV/cm) as shown in figure-6.

It is reported in the literature that by doping at Ba site low ferroelectric phase transition (orthorhombic-rhombohedral $T_{O-R}$ and tetragonal -orthorhombic $T_{T-O}$) shifts to low temperature side and vice versa for the doping at Ti site [5, 6]. A careful look of the obtained data indicates that for the low concentration of Ca doping (Ca2 and Ca5), the above mentioned transition temperature are not shifted significantly. For Ca2 and Ca5 the $T_{O-R}$ are 212 K and 209 K, while $T_{T-O}$ are 273 K and 271 K respectively. However for the Ca10 sample pronounced change in the transition temperature has been observed ($T_{O\text{-}R}$= 200 K, $T_{T\text{-}O}$= 250 K). The other observation is that up to the highest doping of Ca, 10% in the present work, frequency independent behavior has been observed around the transition temperature as mentioned above. For about 10% Sn doping at Ti site in pure BTO, all the three transition temperatures merges into one broad peak and there is direct transition from rhombohedral to cubic structure [12]. However, as observed in the present study even up to 10% Ca doping the transition temperatures ($T_{O\text{-}R}$= 200 K, $T_{T\text{-}O}$= 250 K) are found to be frequency independent and sharp, characteristics of the first order phase transition. The third observation is the enhancement in the magnitude of the dielectric constant at room temperature of the all studied compound as compare to pure BTO. One would like to mention here that at room temperature the dielectric constant of pure BTO is close to 2000 [15, 16], whereas the dielectric constant for Ca2, Ca5 and Ca10 are 3623, 3823 and 2933 respectively at frequency of 50 kHz. Whereas room temperature tanδ (not shown) is in the range of 0.03 for all the studied samples. For Ca2 and Ca5 sample, the transition temperature correspond to the orthorhombic-rhombohedral $T_{O\text{-}R}$ and tetragonal-orthorhombic $T_{T\text{-}O}$ are almost equivalent to pure BTO with enhanced dielectric values. The present results show that it is possible to enhance the dielectric properties of BTO based ceramics with co-doping at Ba and Ti site without modifying/disturbing transition temperature of pure BTO. As a result of better dielectric and ferroelectric properties, the Ca and Sn co-doped BTO ceramics can be considered as suitable for EC based applications and therefore EC coefficients are calculated as discussed below.

To evaluate EC properties we have used indirect method, which is based on the Maxwell relation $(\delta P/\delta T)_E = (\delta S/\delta E)_T$ [1, 2, 3, 4]. For a material having density ρ and heat capacity C, the adiabatic temperature change (ΔT) is defined as

$$\Delta T = -\frac{1}{\rho C}\int_0^E T\left(\frac{\partial P}{\partial T}\right)_E dE \qquad (1)$$

For calculating the EC value, detailed temperature dependent PE loop measurements has been carried out across the ferroelectric to paraelectic transition for all the studied samples. It is desirable to study the electro-caloric effect near to ferroelectric to para electric transition as there is maximum change in entropy across the transition. High temperature P-E data of all studied ceramics has been presented in figure 6. And figure7 shows the variation of maximum polarization as a function of temperature across the ferro-para electric transition for various applied electric field. The $P_{max}$ value has been extracted from the upper branch of the corresponding P-E loop. The value of $\partial P/\partial T$ has been calculated from the variation of $P_{max}$ with the temperature as shown in figure 7. The density are 5.76 g/cc, 5.71 g/cc and 5.65 g/cc for Ca2, Ca5 and Ca10 samples respectively as measured by liquid displacement method. It has been reported in the literature that by use of effective specific heat capacity, there are reasonable good agreement between direct and indirect methods [4]. We have taken effective specific heat values from literature [12] as commonly used in the past literature [4, 17, 18, 19]. For calculating the adiabatic temperature change $dE=E_2-E_1$, has been used, where $E_2$ is the maximum applied electric field and $E_1$ is set to zero as reported in the literature [8, 9, 12]. The adiabatic temperature, ΔT are plotted in the inset of figure7 for all the studied samples. Further the electro-caloric strength, ζ= dT/dE is also calculated across the transition temperature (ferroelectric to paraelectric) and comes out to be 0.25 K mm/kV, 0.49 K mm/kV and 0.27 K mm/kVfor Ca2, Ca5 and Ca10 samples respectively. For Ca5 sample, across the transition temperature (385 K) we achieved giant electro-caloric strength of around 0.49 K mm/kV, which is highest as compared to other BTO based ceramicsas shown in table-1.

## 4. Conclusions

In conclusion, structural, Mössbauer, ferroelectric, dielectric and electro-caloric study is reported on polycrystalline $Ba_{1-x}Ca_xTi_{0.95}Sn_{0.05}O_3$ (x=2, 5 and 10%) ceramics. X-ray diffraction pattern along with Reitveld refinements confirms the single phase behavior of all the studied compound. Iso-valent substitution of Sn has been examined and verified by the $^{119}$Sn Mossbauer spectra. All studied compounds remained tetragonal and ferroelectric at room temperature. Orthorhombic to rhombohedral & tetragonal to orthorhombic ferroelectric transitions shifted to lower temperature side, while cubic to tetragonal shifted to higher temperature side with the increase of Ca content. Enhanced ferroelectric and dielectric properties are observed as compared to pure $BaTiO_3$ at room temperature. Giant electro-caloric strength around 0.5 K mm/kV has been observed for

$Ba_{0.95}Ca_{0.05}Ti_{0.95}Sn_{0.05}O_3$ samples at 385 K.

**Acknowledgments:** Author would like to thank Prof. E. V. Sampathkumaran, TIFR Mumbai (India) for the dielectric data.

**TABLE 1**. Various BTO based ceramics along with their respective electro-caloric strength (ζ).

| Materials (Ceramic) | Tc (K) | ΔT (K) | ΔE (kV/cm) | ζ (K mm/kV) | References |
|---|---|---|---|---|---|
| $Ba_{0.8}Ca_{0.2}Zr_{0.04}Ti_{0.96}O_3$ | 386 | 0.27 | 7.95 | 0.34 | 14 |
| $BaTi_{0.95}Sn_{0.05}O_3$ | 300 | 0.069 | 3.62 | 0.19 | 12 |
| $Ba_{0.94}Ca_{0.06}Ti_{0.875}Sn_{0.125}O_3$ | 298 | 0.24 | 6 | 0.4 | 9 |
| $Ba_{0.92}Ca_{0.08}Ti_{0.92}Zr_{0.08}O_3$ | 320 | 0.38 | 15 | 0.25 | 8 |
| $0.7BaZr_{0.2}Ti_{0.8}O_3$-$0.3Ba_{0.7}Ca_{0.3}TiO_3$ | 328 | 0.3 | 20 | 0.15 | 10 |
| $Ba_{0.65}Sr_{0.35}TiO_3$ | 333 | 0.23 | 10 | 0.23 | 20 |
| $Ba_{0.95}Ca_{0.02}Ti_{0.95}Sn_{0.05}O_3$ | 370 | 0.2 | 7.89 | 0.25 | Present work |
| $Ba_{0.95}Ca_{0.05}Ti_{0.95}Sn_{0.05}O_3$ | 386 | 0.53 | 10.81 | 0.49 | Present work |
| $Ba_{0.95}Ca_{0.10}Ti_{0.95}Sn_{0.05}O_3$ | 398 | 0.34 | 12.69 | 0.27 | Present work |

## Figure captions

**Figure 1.** X-ray diffraction data (along with Reitveld Refinmnet) of $Ba_{1-x}Ca_xTi_{0.95}Sn_{0.05}O_3$ (x=2, 5 and 10%) ceramic at room temperature.

**Figure 2.** SEM image of $Ba_{1-x}Ca_xTi_{0.95}Sn_{0.05}O_3$ (x=2, 5 and 10%) ceramic at room temperature. The scale is 1 µm for all samples.

**Figure 3.** $^{119}$Sn Mössbauer data of $Ba_{1-x}Ca_xTi_{0.95}Sn_{0.05}O_3$ (x=2, 5 and 10%) ceramic at room temperature.

**Figure 4.** PE loop for $Ba_{1-x}Ca_xTi_{0.95}Sn_{0.05}O_3$ (x=2, 5 and 10%) ceramic at room temperature.

**Figure 5.** Dielectric constant data with variable frequency (1 kHz-100 kHz) for $Ba_{1-x}Ca_xTi_{0.95}Sn_{0.05}O_3$ (x=2, 5 and 10%) ceramic as a function of temperature. The solid dashed line is guide to eye.

**Figure 6.** High temperature PE loop for $Ba_{1-x}Ca_xTi_{0.95}Sn_{0.05}O_3$ (x=2, 5 and 10%) ceramic across its ferroelectric to paraelectric phase transition with applied electric field of 7.89 kV/cm, 10.81 kV/cm and 12.69 kV/cm respectively.

**Figure 7.** Maximum polarization ($P_{max}$) as extracted from the high temperature PE data for different applied electric field. Inset shows the corresponding variation in ΔT as a function of temperature.

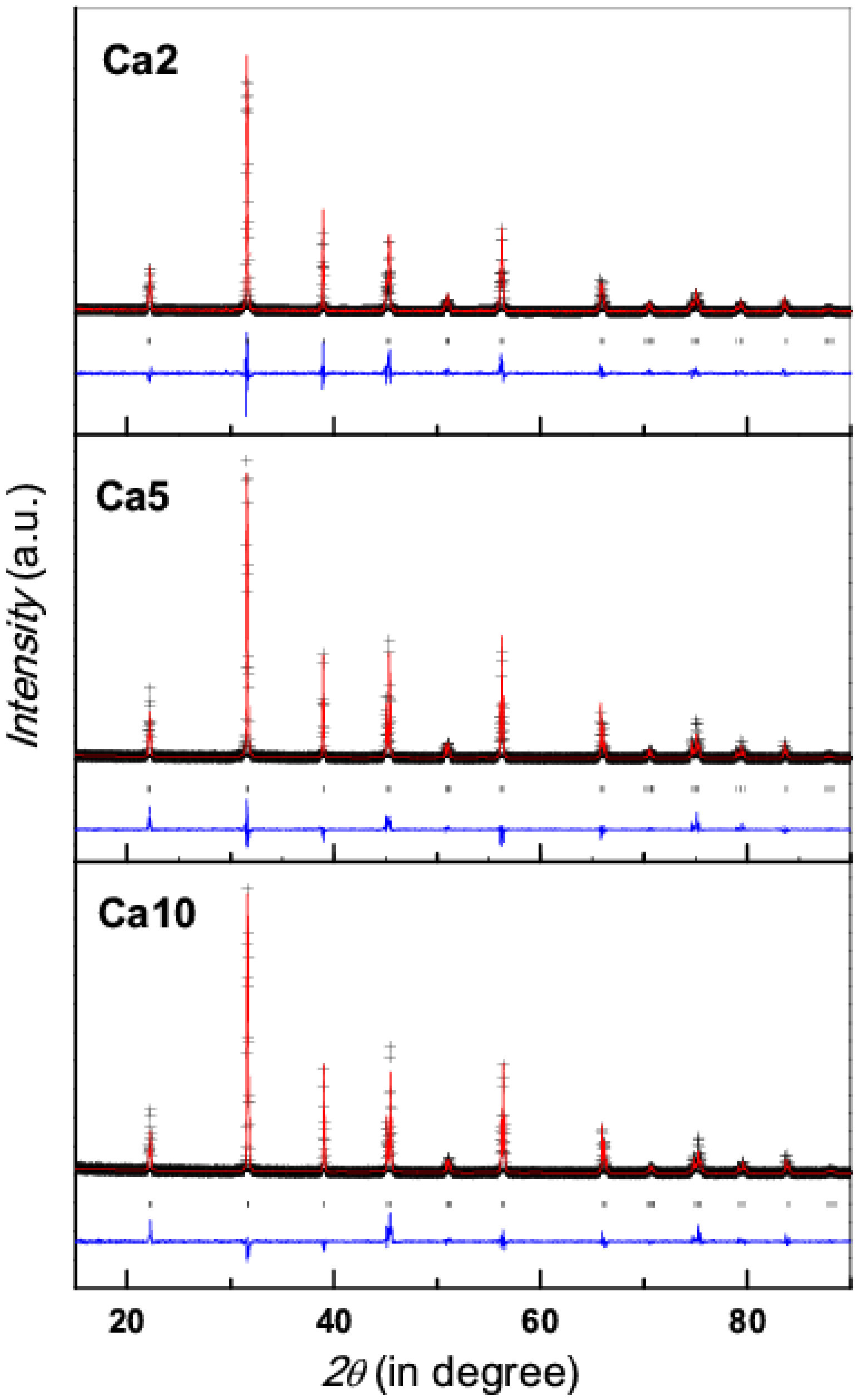

**Figure 1**

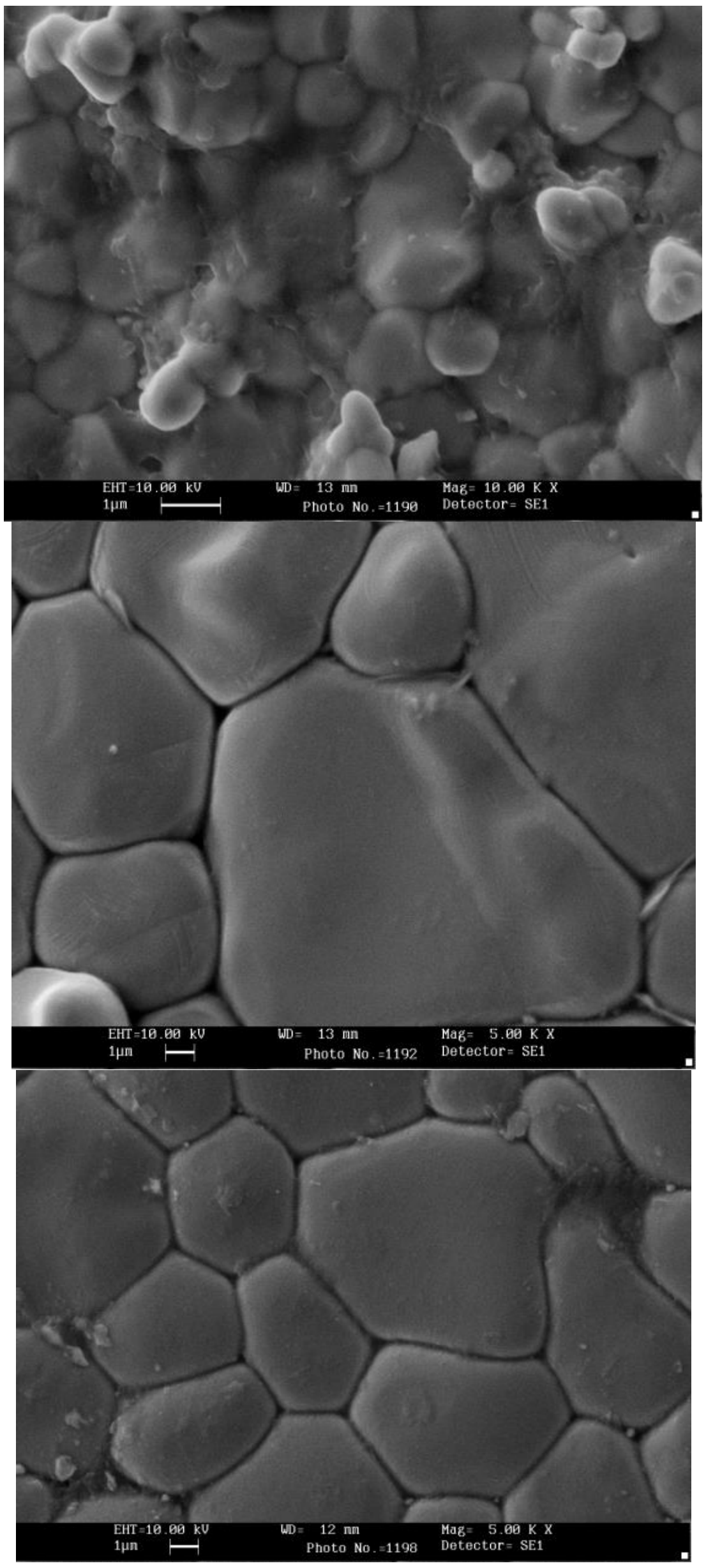

**Figure 2**

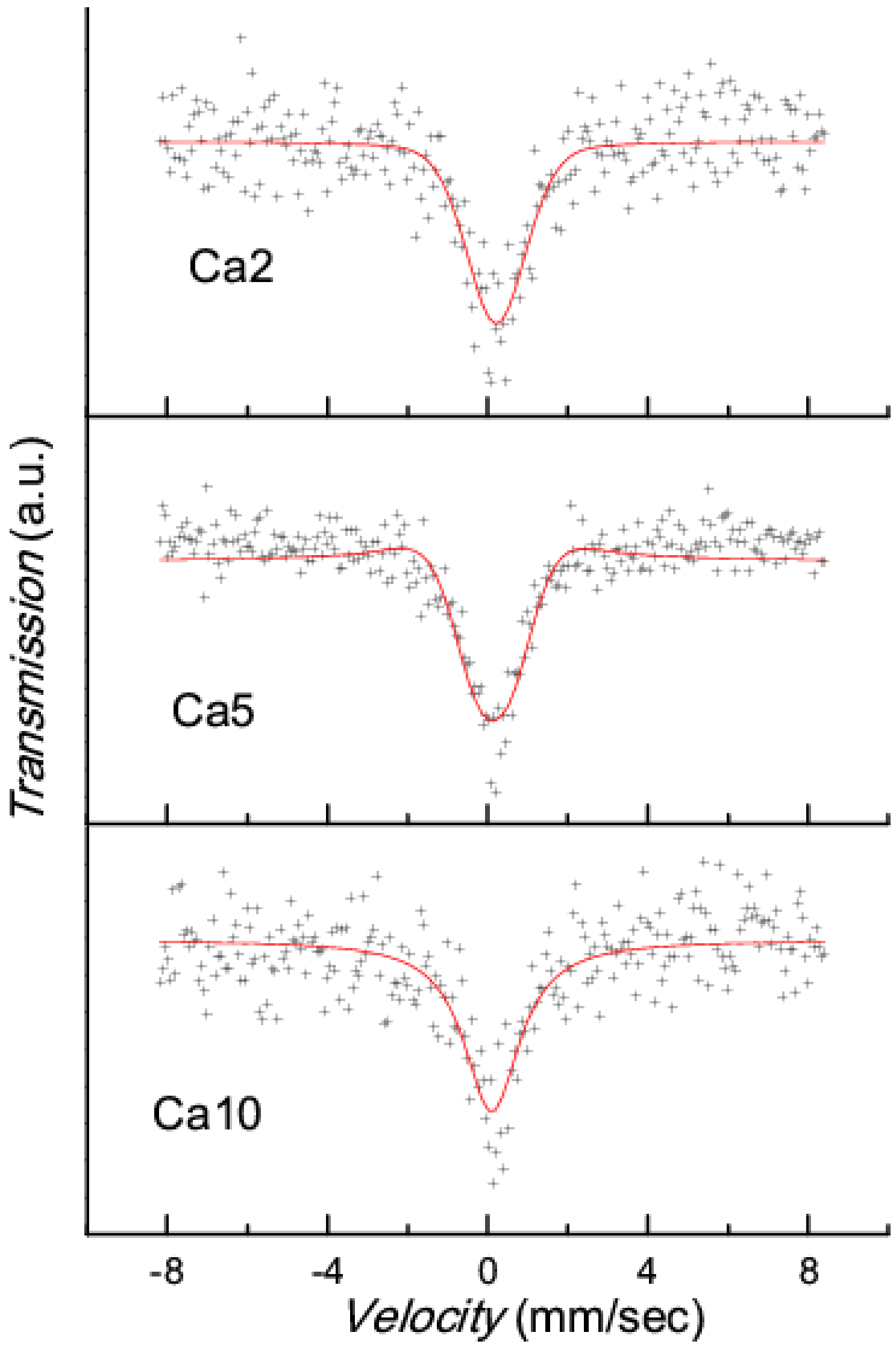

**Figure3**

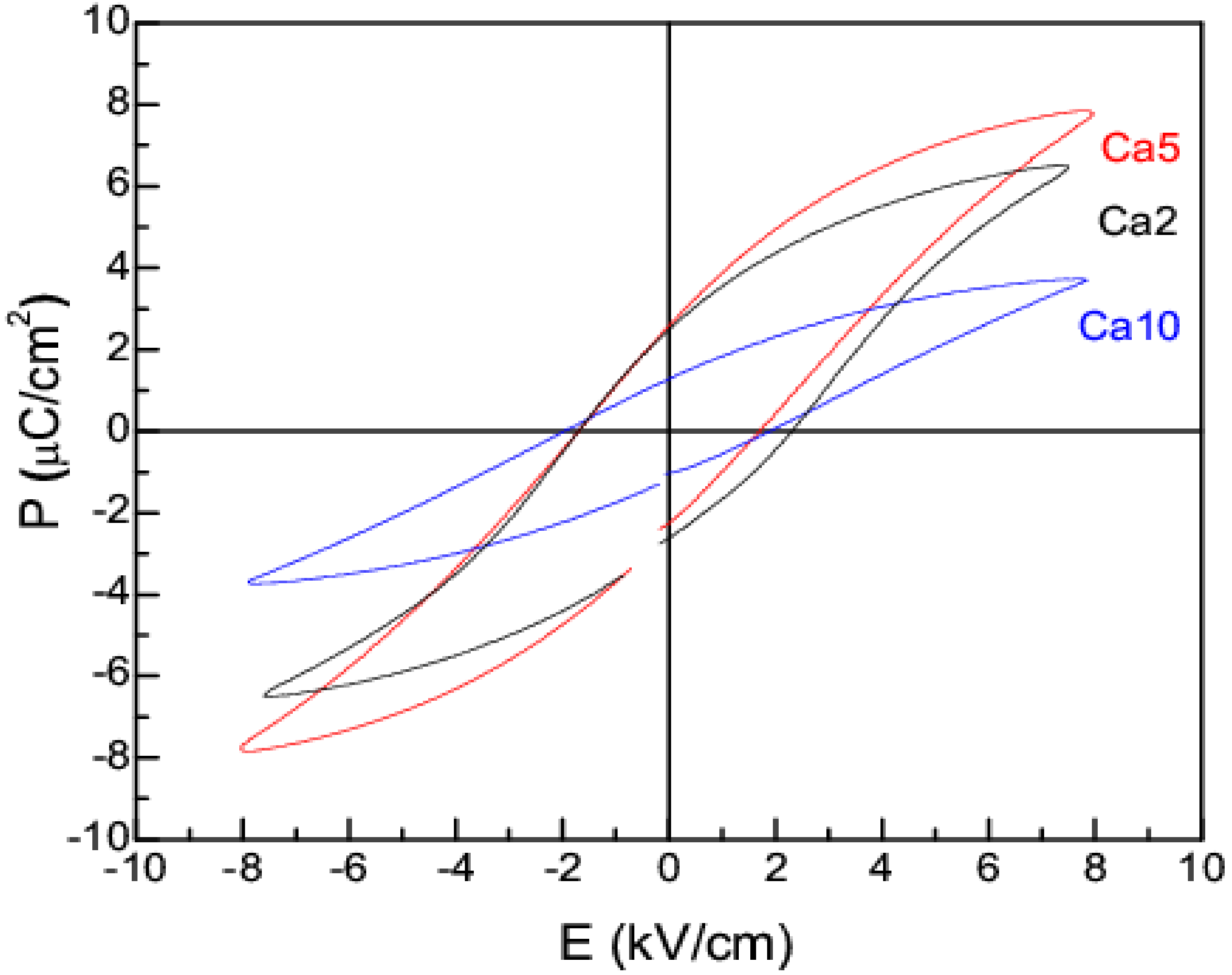

**Figure 4**

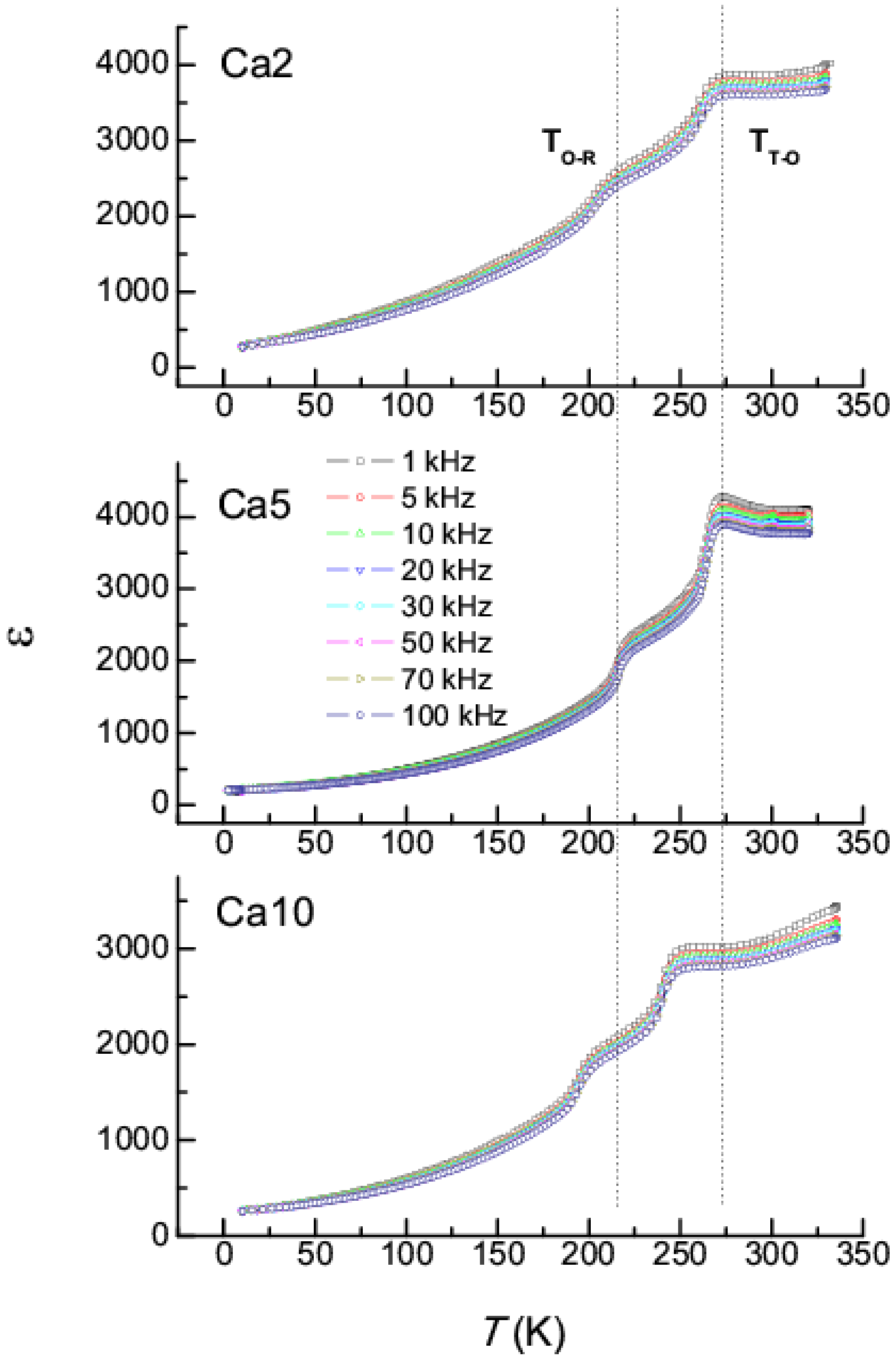

**Figure5**

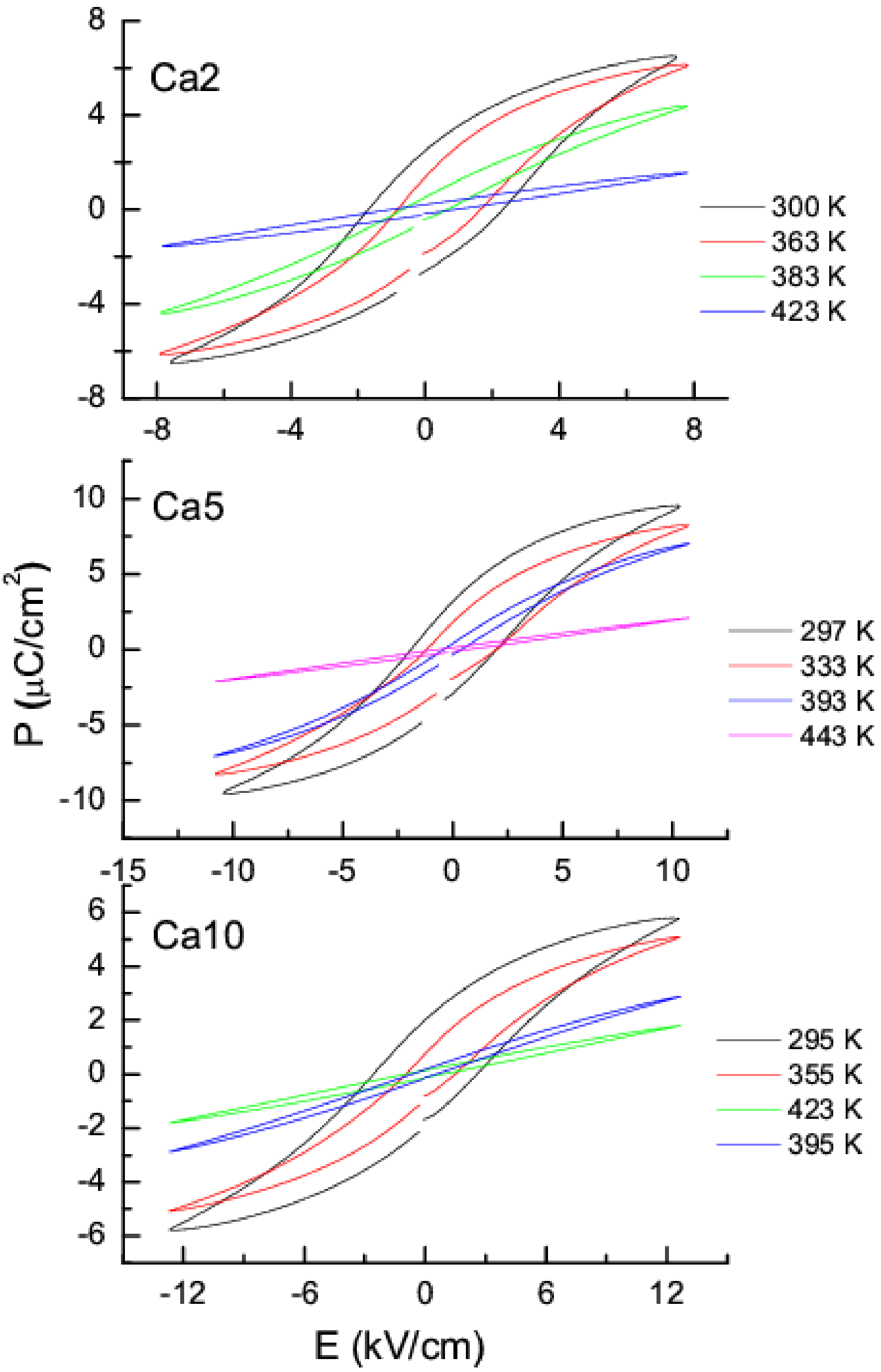

**Figure 6**

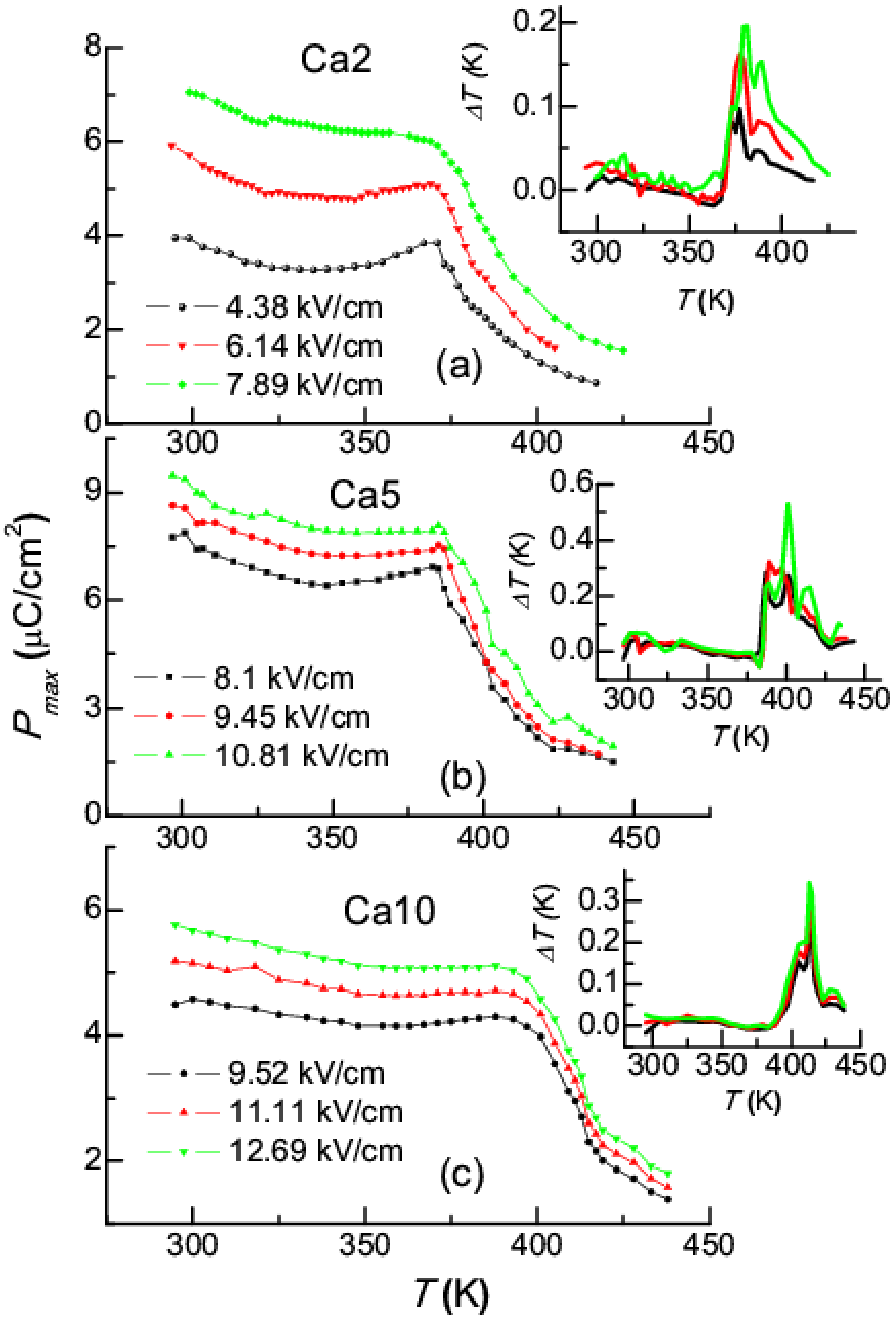

**Figure7**